\documentstyle[twoside,fleqn,espcrc2]{article}

\def\epsfpreprint{Y}   

\if \epsfpreprint Y \input epsf \fi
\def\captionA{$\ E_{eff}(t)$ for the $L=16$, $L_t=16$ lattice.}
\def\captionB{$\ E_{eff}(t)$ for the $L=8$, $L_t=16$ lattice.}

\begin{document}

\title{\bf The $I=1$, $J=1$ channel of the O(4) $\lambda \phi^4_4$
theory.}
\vskip 1. truein
\author{Khalil M. Bitar\thanks {speaker} and Pavlos M. Vranas \\
Supercomputer Computations Research Institute \\
The Florida State University \\
Tallahassee, FL 32306--4052 }

\vskip 1.5 truein

\begin{abstract}

A Monte Carlo simulation of the $O(4)$ $\lambda \phi^4$ theory in the
broken phase is performed on a hypercubic lattice in search of an
$I=1$, $J=1$ resonance. We investigate the region of the cutoff theory
where the interaction is strong as it is there that a resonance would
be expected to have a better chance to form. In that region the
presence of an $I=1$, $J=1$ resonance with mass below the cutoff is
excluded.

\end{abstract}

\maketitle

The simplest effective theory describing low energy pion nucleon
interactions is described by a chiral Langrangian, and involves the
three pion fields $\pi^1, \pi^2, \pi^3$, the scalar field $\sigma$,
the two nucleons, their interactions, and the dimensionful pion decay
constant $f_\pi=90~MeV$. The $\sigma$, although it has never really
been seen, may exist in nature as a broad and quite heavy ($\approx
700~MeV$) isospin I=0, spin J=0 resonance.  In that setting, it was
natural to ask whether the presence of the $\rho$ resonance was a
consequence of the pion interactions or of some more fundamental
interactions.

Since the earliest such theory( Gell--Mann Levy sigma
model~\cite{Gell--Mann&Georgi}) becomes an O(4) $\lambda \phi^4$
theory if the nucleon fields are neglected, this question reduces to
whether or not the O(4) $\lambda \phi^4$ theory can sustain an $I=1$,
$J=1$ resonance in the broken phase.

Today, in a very different setting, this same question is of interest
again.  The scalar sector of the Minimal Standard Model is also a four
component $\lambda \phi^4$ theory in the broken symmetry phase.  The
equivalent of the $\sigma$ resonance is the Higgs particle, the three
pions are the Goldstone bosons, the pion decay constant is the weak
scale $f_G=246~GeV$, and the scattering of longitudinally polarized
vector bosons behaves exactly the same way as a scaled up version of
$\pi$--$\pi$ scattering (equivalence theorem~\cite{equivalence}).

It is possible that the Higgs, like the $\sigma$, is quite heavy and
broad and may avoid detection at SSC.  On the other hand, if the four
component $\lambda \phi^4$ theory does indeed contain an $I=1$, $J=1$
resonance, then since the $\rho$ resonance, as it appears in nature,
is quite strong, its equivalent in the scalar sector of the Minimal
Standard Model may have a good chance to be detected at SSC.
Therefore, it becomes very important to know if another ``signature''
of the scalar sector, besides the Higgs resonance, is awaiting
discovery at SSC.

Earlier attempts~\cite{Basdevant&Lee} to answer this question used
analytical approximations such as Pade approximants and found that an
$I=1$, $J=1$ resonance is present in the theory. It is not clear,
however, whether or not the presence of this resonance is an artifact
of the approximations used and therefore the question of the existence
of an $I=1$, $J=1$ resonance in the theory has yet to receive a
conclusive answer.

We report here on a Monte Carlo simulation to shed new light on this
old question~\cite{KMB&PMV}.  The O(4) $\lambda \phi^4$ theory has the
lattice action
\begin{eqnarray}
S &=& \sum_{x \in \Lambda}\left\{
-k \sum_{\mu=1}^{4} \left( \vec\Phi_x\vec\Phi_{x+\hat\mu}+
\vec\Phi_x\vec\Phi_{x-\hat\mu} \right) +\nonumber \right. \\
& & \left.\lambda (\vec\Phi_x\vec\Phi_{x}-1)^2+
\vec\Phi_x\vec\Phi_{x}
\right\} \label{action}
\end{eqnarray}
and was simulated on the lattice in the $\lambda \rightarrow \infty$
limit. In that limit, the theory has the strongest interactions and a
resonance probably has a better chance to form.

The simulation was done on a hypercubic lattice $\Lambda$ of spatial
extension $L$ and time extension $L_t$ using an incomplete heat bath
algorithm~\cite{Fredenhagen&Marcu} on the CM-2 machine at SCRI.
Since, on a finite lattice, the direction of the symmetry breaking
changes from configuration to configuration, the same approach as
in~\cite{Has} was used to disentangle the massive scalar field
$\sigma$ from the Goldstone modes.

\addtocounter{footnote}{-1}
The mass $m_\sigma$ of the $\sigma$ field was measured using standard
techniques. To measure the energy of the lowest laying state in the
$I=1$, $J=1$ channel, an operator carrying these quantum numbers needs
to be constructed. The simplest such operator is:
\begin{equation}
O_{c,m}(t)={1\over{L^3}} \sum_{x \in \Lambda_t} \pi_x^a \pi_{x+\hat
m}^b \epsilon_{abc}
\end{equation}
where summation over repeated
indices is assumed, $a, b, c $ are the isospin indices,
$\epsilon_{abc}$ is the totally antisymmetric tensor, $\hat m$ is the
$m$'th Euclidean unit vector of the time slice $\Lambda_t$, and $m \in
[1,2,3]$ is the z-component spin index.  Unfortunately, the time slice
connected correlation function of this operator gives a very weak
signal.  To get a better signal an operator that extends over several
lattice spacings needs to be constructed using a trial wave function
for the two--pion state. The ``bag'' and ``bound state'' meson wave
functions were considered~\cite{DeGrand&Loft}. The latter gave a
better signal with an affordable cost in computer time and was
therefore used for the simulation.  Using as a trial wave function for
the two pions at relative position $\vec R$, the hydrogen wave
function $\Psi_{n=2,l=1,m}(\vec R)$, an $I=1$, $J=1$ operator with
total 3-momentum zero was constructed:
\begin{eqnarray}
& & O_{c,m}(t)=\sum_{x \in \Lambda_t} \sum_{\vec R \in B} \nonumber \\
& & |\vec R| \exp({-|\vec R|/2a_0})
Y_{1,m}(\theta,\phi) \pi_x^a \pi_{x+\vec R}^b \epsilon_{abc}
\label{rho_op}
\end{eqnarray}
where $B$ is a three-dimensional cubic box centered at the origin and
contained in $\Lambda_t$, $a_0$ is the ``Bohr radius'' in lattice
units, $\theta$ and $\phi$ are the azimuthal and polar angles of $\vec
R$, and $Y_{1,m}$ is the $l=1$ spherical harmonic (up to a
multiplicative normalization constant).  The parameter $a_0$ can be
given any value. A very large $a_0$ will cause the exponential to
decrease very slowly and then the sum over $\vec R$ will have to be
carried out over a box $B$ as large as $\Lambda_t$.  Since this can
increase the computer time significantly, a smaller $a_0$ has to be
used so that the size of the box that contains the important
contribution from the exponential can be made smaller. However, $a_0$
cannot be made very small because the signal becomes weaker as $a_0$
decreases. An optimal choice of $a_0$ and $B$ was found to be $a_0=2$,
and $B$ extending from $-3$ to $+3$ in each of the three directions.
The operator $O_{c,m}(t)$ couples to the $I=1$, $J=1$ states.  The
energy of the lowest laying state in this channel can be found by
looking at the time slice connected correlation function
$C_{c,m}(t)=<O_{c,m}(0)O_{c,m}(t)^*>_c$ of this operator.

The simulation was done on an $L=8, L_t=16$ and $L=16, L_t=16$ lattice
and for three values of the hopping parameter $\kappa=0.305, 0.310,
0.330$. These values were chosen so that a wide range of $m_\sigma$
will be covered (they also coincide with some of the values used
in~\cite{Has}). The expectation values of $C_{1,1}(t)={1\over
9}\sum_{c,m} C_{c,m}(t)$, were measured.
\begin{table}
\begin{center}
\begin{tabular}{||c|c|l|c|c||}  \hline
$\kappa$ & L  & $\ \ \ E$ & fit--range & $\chi^2/$d.o.f.\\ \hline\hline
0.305    & 16 & 0.94(1)   &   2--8     &     8.8        \\ \hline
0.305    & 16 & 0.76(3)   &   3--8     &     1.8        \\ \hline
0.310    & 16 & 0.94(1)   &   2--8     &     0.7        \\ \hline
0.310    & 16 & 0.91(3)   &   3--8     &     0.7        \\ \hline
0.330    & 16 & 0.94(1)   &   2--8     &     4.2        \\ \hline
0.330    & 16 & 0.87(2)   &   3--8     &     2.4        \\ \hline
0.305    &  8 & 1.549(5)  &   1--5     &     1.7        \\ \hline
0.310    &  8 & 1.531(6)  &   1--5     &     1.7        \\ \hline
0.330    &  8 & 1.501(9)  &   1--4     &     0.4        \\ \hline
\end{tabular}
\end{center}
\caption{Energy $E$ in the $I=1$, $J=1$ channel.}
\end{table}
$C_{1,1}(t)$ was fitted for a few different ranges of $t$, and the
resulting energies $E$, together with the $\chi^2$ per degree of
freedom for each fit are given in table 1 for the $L=16$ and $L=8$
lattices.

The effective energy $E_{eff}(t)$, for the two time slices at t-1 and
t, is plotted versus t for the three values of $\kappa$ in figures 1
($L=16$) and 2 ($L=8$). The values of t omitted from those plots had
an $E_{eff}(t)$ with very large error.

In a two-pion $I=1$, $J=1$ state with zero total 3--momentum, the
lowest 3--momentum a pion can have has one component equal to
$2\pi\over L$ and two equal to $0$.  The next one has two components
equal to $2\pi\over L$ and one equal to $0$. The energy spectrum in
the $I=1$, $J=1$ channel is therefore expected to contain levels with
energies close to the energies of these states, but slightly different
because of the interaction. For the $L=16$ lattice, these levels have
energies $E_0\simeq 0.78$ and $E_1\simeq 1.09$ respectively, and are
denoted by the dotted lines
\if \epsfpreprint Y
\begin{figure}[t]
\epsfxsize=\columnwidth
\epsffile{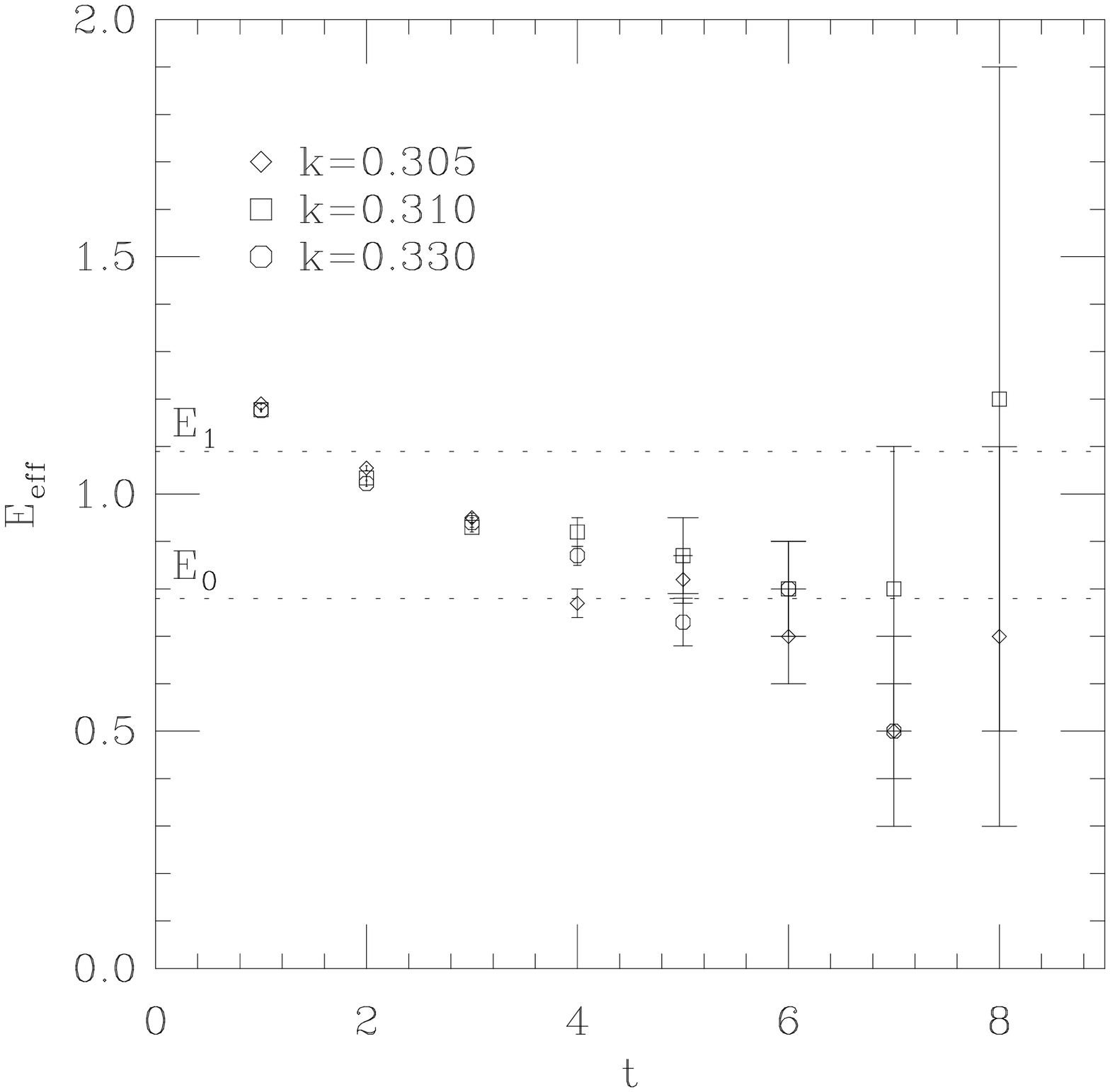}
\caption{\captionA}
\end{figure}
\fi
in figure 1.  From this figure, it is clear that the observed energy
levels are very close to the levels of two free pions.  In fact, for
smaller $t$, the levels are close to $E_1$, and for larger $t$ they
are close to $E_0$.  Because the free two--pion levels are not very
well separated at $L=16$, the observed levels are probably a mixture
of the two lowest ones. In that sense, the energies in table 1 for the
$L=16$ lattice are probably a mixture as well. The fact that the
observed levels correspond to a two--pion state and not to a resonance
is also greatly supported by the fact that these levels change only
slightly from $\kappa=0.305$ to $\kappa=0.330$.  After all, in that
range $m_\sigma$ varies from $0.225$ to $0.91$. Therefore, if a
resonance is present it must have energy larger than $\approx 0.78$
and is either too heavy (for example, heavier than $1.09$) to be
observed with this method,
\if \epsfpreprint Y
\begin{figure}[t]
\epsfxsize=\columnwidth
\epsffile{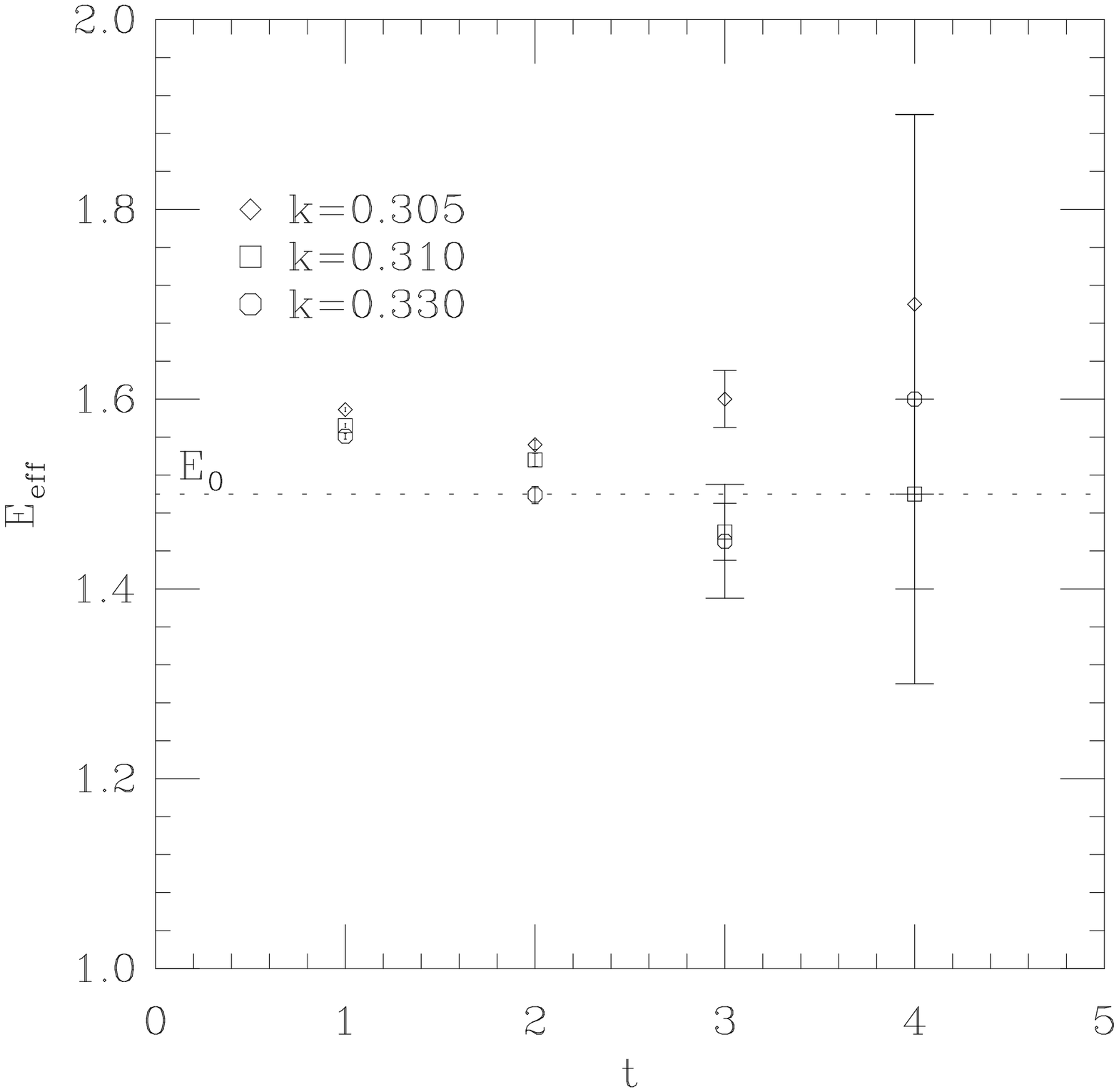}
\caption{\captionB}
\end{figure}
\fi
or is ``hiding'' between $0.78$ and $1.09$. That the latter is not the
case can be seen by looking at the energy levels obtained for the
$L=8$ lattice. From figure 2 it is clear that there are no
energy levels below $\approx 1.4$. In fact, since the two lowest free
two--pion states are well separated in this lattice, the lowest energy
level is clearly visible in this figure .  It is true that the limited
statistics give $E_{eff}(t)$ only up to $t=4$, but because the lowest
level is well separated from the next one, the correlated $\chi^2$ fit
gives a good estimate of the energy of this level. The numbers given
in table 1 are indeed very close to the lowest free two--pion energy
$E_0\simeq 1.50$ and do not seem to change much for the different
values of $\kappa$.

{}From this analysis it is evident that if an $I=1$, $J=1$ resonance
exists for $\kappa \ge 0.305$, it is unphysical since it must have
energy larger than the cutoff. Of course there is always the
possibility that a resonance with energy below the cutoff does exist,
but, because the overlap of the operator in eq. $3$ with an $I=$,
$J=1$ resonance may have been much smaller than its overlap with the
free two pion states present in that channel, it was not possible to
observe with the statistics and lattices used.  However, it is
unlikely that this is the case since the operator in eq. $3$ is more
appropriate for a resonance than for a free two particle state.

For the theory of Gell-Mann and Levy of low energy pion processes the
$\kappa \ge 0.305$ region corresponds to the region where the $\sigma$
particle mass is greater than approximately $180~MeV$ (or equivalently
the cutoff is less than approximately $1.3~GeV)$.  \footnote{ The
connection with the physical scale $f_\pi$ (or $f_G$) is made through
the renormalized coupling constant $g_R=3m_\sigma^2/f_\pi^2$. The
value of $g_R$ for this value of $\kappa$ was taken
from~\cite{Luscher&Weisz}.} Since $m_\sigma$ is expected to be much
larger than $180~MeV$, it is clear from the results that the existence
of the $\rho$ resonance in nature cannot be accounted for by this low
energy theory alone (in contrast to~\cite{Basdevant&Lee}).

For the scalar sector of the Minimal Standard Model the $\kappa \ge
0.305$ region corresponds to where the Higgs mass is greater than
approximately $500~GeV$ (or equivalently the cutoff is less than
approximately $3.5~TeV$). The Higgs mass is of course not known but
its upper bound is placed at around $650~GeV$ (hypercubic lattice
triviality bound). Thus the existence of an $I=1$, $J=1$ resonance can
be excluded with confidence for values of the Higgs mass above
$\approx 500~GeV$. For $\kappa< 0.305$ (Higgs mass $< 500~GeV$) the
strength of the interaction becomes weaker and hence it is safe to say
that if a resonance could not form for $\kappa \ge 0.305$, where the
interaction is stronger, it is unlikely that it will in this region
either. For this reson it was not deemed necessary to investigate the
$\kappa< 0.305$ region. A numerical simulation for $\kappa< 0.305$ not
only is not necessary, but it would also be very costly since larger
lattices will have to be used (the correlation length becomes larger
than approximately ${1\over 0.225}\simeq 4.5$).

It must be emphasized that these conclusions are valid only within the
realm of the scalar sector of the Minimal Standard Model.  It is of
course still possible that the physics that enters at higher energies
may be able to produce such a resonance in very much the same way the
physical $\rho$ particle owes its existence to QCD.  This resonance if
it exists due to some higher energy theory it would have energy
determined by that theory. In fact, it is possible that the energy of
this resonance is determining the cutoff energy of the Minimal
Standard Model.

\bigskip

\if \epsfpreprint N
\medskip
\section*{FIGURE CAPTIONS}
\smallskip

\noindent{\bf Figure~$1$:~}\captionA

\noindent{\bf Figure~$2$:~}\captionB
\fi

\section*{ACKNOWLEDGMENTS}

This research was partially funded by the U.S. Department of Energy
through Contract No. DE-FG05-92ER40742 and DE-FC05-85ER250000.

\vfill


\begin{thebibliography}{9}
\bibitem{Gell--Mann&Georgi} M. Gell--Mann, M. Levy, {\bf Il Nuovo
Cimento Vol. XVI, N. 4} (1960) 705; H. Georgi, {\bf Weak
Interactions and Modern Particle Theory} (1984) Addison-Wesley Pub.
\bibitem{equivalence} M.S. Chanowitz, M.K. Gaillard
{\bf Nucl. Phys.  B261} (1985) 379.
\bibitem{Basdevant&Lee} See for example: J.L. Basdevant, B.W. Lee
{\bf Phys. Rev. D Vol. 2, No 8} (1970) 1680; LH. Chan, R.W. Haymaker {\bf
Phys. Rev. D Vol. 10, No. 12} (1974) 4170.
\bibitem{KMB&PMV} K.M. Bitar, P.M. Vranas {\bf Phys. Lett. B 284}
(1992) 366.
\bibitem{Fredenhagen&Marcu} K.
Fredenhagen, M. Marcu {\bf Phys. Lett.  B 193} (1987) 486.
\bibitem{Has} A. Hasenfratz, K. Jansen, J. Jersak, C.B. Lang,
T. Neuhaus, H. Yoneyama {\bf Nucl. Phys. B317} (1989) 81.
\bibitem{DeGrand&Loft} T.A. DeGrand, R.D. Loft {\bf Comp. Phys.
Comm.  65} (1991) 84.
\bibitem{Luscher&Weisz} M. L\"uscher, P. Weisz
{\bf Nucl.  Phys. B318} (1989) 705.
\end{thebibliography}
\end{document}